## Perpendicular Ion Heating by Cyclotron Resonant Dissipation of Turbulently Generated Kinetic Alfvén Waves in the Solar Wind

Philip A. Isenberg and Bernard J. Vasquez

Space Science Center and Department of Physics,

University of New Hampshire,

Durham, NH 03824 USA

**Abstract.** Solar wind ions are observed to be heated in the directions perpendicular to the large-scale magnetic field, with preferential heating given to heavy ions. In the solar corona, this heating may be responsible for the generation of the wind itself. It is likely that this heating results from the dissipation of plasma turbulence, but the specific kinetic mechanism that produces these preferential effects is not known. Solar wind turbulence at proton scales is often characterized in terms of highly oblique kinetic Alfvén waves (KAWs), which have been thought to dissipate through the Landau resonance and yield parallel heating. We show that the quasilinear resonant cyclotron interaction between KAWs and solar wind ions can actually produce perpendicular ion heating. We present an illustrative calculation of a steady, critically balanced spectrum of KAWs acting on homogeneous ion distributions with a plasma $\beta = 0.1$, representative of turbulent conditions in the fast solar wind near 20 $R_s$. The KAWs are described here by a two-fluid dispersion relation. We find that thermal protons are strongly heated in the perpendicular direction within a typical quasilinear time of several thousand gyroperiods, which corresponds to only a few minutes at 20 $R_s$. Alpha particles in the same fluctuation field are heated to similar perpendicular thermal speeds, equivalent to the greater than mass proportional perpendicular temperatures that are commonly observed. We discuss improvements to this simple model that may determine whether this mechanism can be responsible for the observed coronal and solar wind ion heating.

## 1. Introduction

Solar wind ions are observed to be continuously heated in the directions perpendicular to the large-scale magnetic field (Marsch 1991, 2006, 2012; Schwartz & Marsch 1983; Hellinger et al. 2013), with a preferential energization given to ions more massive than protons (Bochsler et al. 1985; von Steiger et al. 1995; Collier et al. 1996; Hefti et al. 1998; Tracy et al. 2016). This preferential perpendicular heating also appears to take place in the corona (Kohl et al. 1998; Esser et al. 1999; Cranmer et al. 2008), and may be responsible for the generation of the solar wind itself (Cranmer & van Ballegooijen 2003; Cranmer 2009; Isenberg & Vasquez 2011, 2015). The specific kinetic mechanism that produces this heating is a subject of vigorous investigation.

The energy for this ongoing heating must be provided by a spatially distributed source, and this requirement points to the dissipation of ubiquitous turbulence in the essentially collisionless plasma of the corona and *in situ* solar wind (Matthaeus et al. 1999; Cranmer & van Ballegooijen 2005; Cranmer et al. 2007; Verdini & Velli 2007; Verdini et al. 2009; Verdini et al. 2010; Schekochihin et al. 2009; Perez & Chandran 2013; Lionello et al. 2014). Solar wind turbulence arises from counter-propagating Alfvénic fluctuations that nonlinearly cascade to smaller scales, where dissipation processes ultimately transfer the fluctuation energy to thermal energy of the plasma particles. This turbulent cascade is known to proceed anisotropically in wavevector space, such that the energy flow is primarily to small-scale fluctuations with high $k_\perp$, where the subscripts $\perp$ and $\parallel$ refer to the directions perpendicular and parallel to the large-scale magnetic field, respectively (Montgomery & Turner 1981; Shebalin et al. 1983; Higdon 1984; Oughton et al. 1994; Goldreich & Sridhar 1995; Cho & Vishniac 2000; Cho et al. 2002).

As this turbulent cascade proceeds to small scales, it is thought to evolve into fluctuations that can be described in terms of highly oblique kinetic Alfvén waves (KAWs) (Leamon et al. 1998; Leamon et al. 1999; Howes et al. 2008a; Howes et al. 2008b; Howes & Quataert 2010; Schekochihin et al. 2009; Bian et al. 2010; Boldyrev & Perez 2012; Boldyrev et al. 2013; Sahraoui et al. 2012; Grošelj et al. 2018a; Arzamasskiy et al. 2019). These turbulent fluctuations are not "waves" in the standard sense, since strong turbulence by definition disrupts the phase of these oscillations within only a few wavelengths. Still, the linear relationships between the particles and the fluctuating fields can be characterized by a dispersion relation, from which polarization and propagation properties can be obtained. A growing number of observational studies in the solar wind and the magnetosheath are found to be consistent with this interpretation (Leamon et al. 1998; Leamon et al. 1999; Bale et al. 2005; Sahraoui et al. 2009; Sahraoui et al. 2010; He et al. 2011; He et al. 2012; Podesta & Gary 2011; Salem et al. 2012; Podesta & TenBarge 2012; Podesta 2013; Chen et al. 2013; Roberts et al. 2013; Klein et al. 2014; Wu et al. 2019).

The validity of this wave-like interpretation is presently being debated (Dmitruk et al. 2001; Narita et al. 2011; Matthaeus et al. 2014; Grošelj et al. 2018b). In this paper, we will proceed as though the primary small-scale interactions between the particles and fields in coronal and solar wind turbulence can be usefully described in terms of randomly phased "quasi-modes" which follow linear dispersion relations. This viewpoint will allow us to investigate the dissipation of these turbulent fluctuations using the formalism of quasilinear (QL) theory. This approach may appear similar to the so-called "quasilinear premise" (Schekochihin et al. 2009; Klein et al. 2012; Howes et al. 2014), which posits that the fundamental properties of plasma turbulence can be understood as a superposition of these linear quasi-modes that are further processed by nonlinear interactions. However, in this paper we do not consider the development of the turbulence or the operation of the nonlinear cascade. Here, we simply address the plausible interaction of a turbulently maintained spectrum of KAW quasi-modes with the ions in the inner heliosphere.

KAWs are Alfvénic fluctuations with perpendicular scales on the order of the proton inertial length, $k_\perp \gtrsim \lambda^{-1}$ or the proton gyroscale, $k_\perp \gtrsim \rho_p^{-1}$, where $\lambda = V_A/\Omega_p$ and $\rho_p = \sqrt{k_B T_p / m_p} / \Omega_p$. Here, $V_A$ is the Alfvén speed in the large-scale magnetic field $B_o$, $V_A = B_o(4\pi m_p n_p)^{-1/2}$, and $\Omega_p$ is the proton gyrofrequency, $\Omega_p = q_p B_o/m_p c$, where $q_p$ and $m_p$ are the proton charge and mass, respectively, $n_p$ is the proton number density, $T_p$ is a characteristic proton temperature, $k_B$ is Boltzmann's constant, and $c$ is the speed of light. The small fluctuation scale means that protons and electrons must be treated separately and the plasma does not behave as a single fluid. Highly oblique

KAWs ($k_\perp/k_\parallel \gg 1$) are compressive ($\delta B_\parallel \neq 0$) and elliptically polarized, with the transverse field fluctuation vectors generally rotating in the right-hand sense (Gary 1986; Hollweg 1999).

Turbulent fluctuations in the regions around these proton scales are small amplitude ($\delta B/B_o \ll 1$), so the linear dissipation of these modes might be expected to describe the particle heating. However, the linear dissipation of KAWs in a thermal plasma is dominated by the Landau resonance, interacting with particles whose parallel speeds match the parallel phase speed of the wave, $\upsilon_\parallel = V_{ph} \equiv \omega/k_\parallel$ (Gary & Borovsky 2004; Gary & Nishimura 2004; Gary & Borovsky 2008; Howes et al. 2008a). Dissipation through the Landau resonance only heats particles with high $\upsilon_\parallel$, and preferentially heats them in the parallel direction, leading to the conclusion that the effective mechanism for the observed perpendicular ion heating in the solar wind must therefore be a nonlinear process. Suggested perpendicular heating mechanisms include turbulent reconnection (Matthaeus & Lamkin 1986; Leamon et al. 2000; Rappazzo et al. 2008; Matthaeus & Velli 2011; Karimabadi et al. 2013), and nonlinear magnetic-moment-breaking (or "stochastic heating") (McChesney et al. 1987; Johnson & Cheng 2001; Chen et al. 2001; White et al. 2002; Chandran et al. 2010; Klein & Chandran 2016; Isenberg et al. 2019).

In this paper, we take a closer look at KAWs and their quasilinear interaction with ions. We note that right-hand, elliptically polarized KAWs still possess a left-hand component, which can resonate with the cyclotron motion of thermal ions. In addition, when the parallel ion plasma $\beta_\parallel = (k_B T_\parallel/m_p)\, V_A^{-2} < 1$, as is often the case in the corona and inner solar wind, most of the thermal ions do not reach the Landau resonance. This last point has been interpreted to mean that protons are not substantially heated by KAWs (Cranmer & van Ballegooijen 2003; Howes et al. 2008a; Howes & Quataert 2010; Howes **2010,** 2011), but we suggest that the residual cyclotron interaction with the core ion distribution may become important for low-$\beta$ plasmas and contribute to the observed perpendicular heating. We will demonstrate the possibility of such heating, using an idealized example of a homogeneous plasma where a steady spectrum of critically balanced KAWs follows a two-fluid dispersion relation.

In the next section, we briefly present the relevant details of the two-fluid dispersion of KAWs, and how these properties may lead to cyclotron-resonant ion heating. In Section 3, we describe the quasilinear analysis of the interaction between thermal ions and KAWs. We apply this analysis, using the two-fluid dispersion relations, to our idealized example, setting the plasma $\beta = 0.1$ and calculating the diffusion coefficients due to a steady spectrum of turbulent KAWs. In Section 4a, we compute the evolution of a proton distribution under this diffusion, starting with a $\beta = 0.1$ Maxwellian. In Section 4b, we consider the QL diffusive heating of alpha particles in the same turbulent fluctuations. In Section 5, we discuss the construction of a more realistic model for this turbulent heating and list some of the implications of this process in the solar wind. Section 6 contains our summary and conclusions.

## 2. KAW Dispersion Relation

In this paper, we will apply a form of the KAW dispersion relation derived from linear two-fluid analysis (Lysak & Lotko 1996; Hollweg 1999; Zhao et al. 2014). This analysis does not include the kinetic effects of the plasma particles, so the self-consistent damping of the fluctuations that would accompany particle heating is not taken into account. Correspondingly, resonant features that may distort the dispersion curves near the ion cyclotron frequencies will not appear in this treatment. The advantage of this approximation is that the solutions are well defined and analytic.

The two-fluid analysis also excludes the effect of coupling between the KAWs and Bernstein waves at the proton cyclotron frequency (Howes et al. 2008a; Podesta 2012). We note that the distortions due to this coupling only affect a narrow frequency band around the proton cyclotron frequency. Furthermore, the Bernstein waves themselves are almost non-propagating, so will not make an important contribution to the energization processes to be discussed here. A recent investigation of fluctuation properties in magnetosheath turbulence has found that the observations are more closely matched by a two-fluid description than by an equivalent kinetic dispersion (Wu et al. 2019).

Since the two-fluid equations and the linearization procedure are well known (e.g. Stix 1992) and the computational details are identical to those in Zhao et al. (2014), we simply present the relevant results for our idealized example. The full two-fluid dispersion relation, valid for the small scales we consider, is given by equation (A17) of Zhao et al. (2014) as a cubic equation for $\omega^2\,(k_\parallel, k_\perp)$. Here, we follow Zhao et al. (2014) further and simplify this relation for $k_\perp >> k_\parallel$ and small electron mass $m_e/m_p << 1$ to give a quadratic expression whose solution is

$$\omega^2 = \frac{k_\parallel{}^2 V_A{}^2 \left(1 + 2\beta + \rho_p{}^2 k_\perp{}^2\right)}{2\left[1 + \lambda_e{}^2 k_\perp{}^2 + \lambda^2 k_\parallel{}^2 + \left(1 + \lambda_e{}^2 k_\perp{}^2\right)^2 \beta\right]}$$

$$\times \left[1 - \sqrt{1 - 4\beta \frac{1 + \lambda_e{}^2 k_\perp{}^2 + \lambda^2 k_\parallel{}^2 + \left(1 + \lambda_e{}^2 k_\perp{}^2\right)^2 \beta}{\left(1 + 2\beta + \rho_p{}^2 k_\perp{}^2\right)^2}}\right], \quad (1)$$

where the electron inertial length $\lambda_e = (m_e/m_p)^{1/2}\lambda$. We also use the KAW polarization properties as given by Equation (8) of Zhao et al. (2014). Taking the electron and proton temperatures to be equal, these are

$$\delta\mathbf{E} = \left[\frac{\omega}{2k_\parallel}\left(\lambda_e{}^2 k_\perp{}^2 + \Gamma_2 + 1\right)\hat{\mathbf{x}} + i\frac{\Omega}{\omega}\Gamma_1\hat{\mathbf{y}} + \frac{\omega}{2k_\perp}\left(\lambda_e{}^2 k_\perp{}^2 + \Gamma_2 - 1\right)\hat{\mathbf{z}}\right]\delta B_y \quad (2)$$

for the Fourier components of the electric field in a coordinate system where the $z$-axis is aligned with the large-scale magnetic field and the wavevector is defined in the $x$-$z$ plane. The $\Gamma$ expressions in (2) are

$$\Gamma_1 = \left(1 + \lambda_e{}^2 k^2\right)\frac{\omega^2}{k^2 V_A{}^2} - \frac{k_\parallel{}^2}{k^2} \quad (3)$$

$$\Gamma_2 = \left(1 + \lambda_e{}^2 k^2\right)^2 \frac{\omega^2}{k^2 V_A{}^2} - \frac{k_\parallel{}^2}{k^2} + \frac{k_\parallel{}^2 V_A{}^2}{\omega^2} - \lambda^2 k_\parallel{}^2 \quad (4)$$

where $k^2 = k_\parallel{}^2 + k_\perp{}^2$.

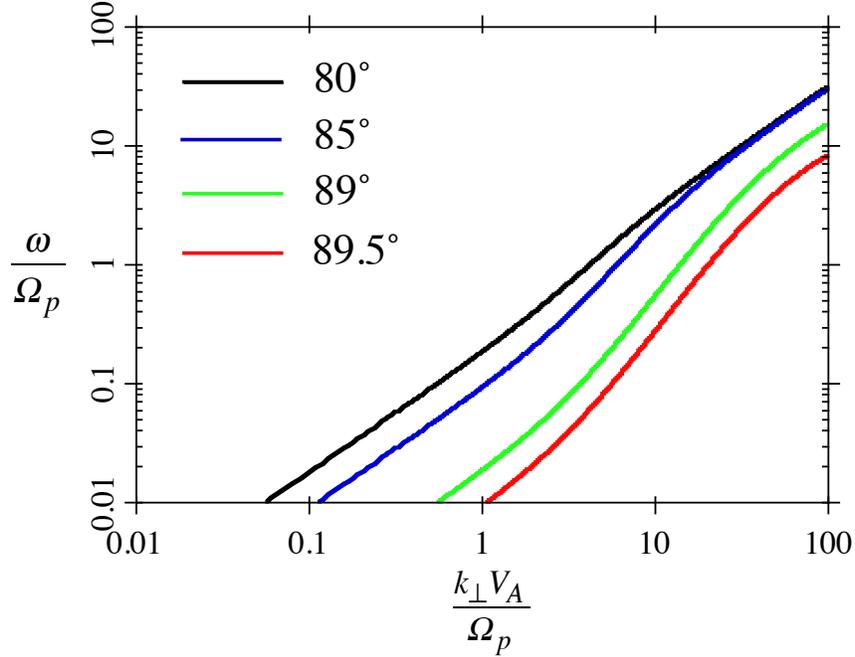

**Figure 1.** KAW dispersion relation from two-fluid analysis for $\beta$ = 0.1 and equal temperatures for protons and electrons. Four propagation angles with respect to the large-scale magnetic field direction are shown (after Zhao et al. 2014).

Figure 1 shows solutions of (1) for four quasi-perpendicular propagation angles when $\beta$ = 0.1. We see that the dispersion curves in this case reach the proton cyclotron frequency and pass smoothly through to higher frequencies. This property implies that the entire proton core can be cyclotron-resonant with these oblique fluctuations.

This resonance with thermal particles is made clear from the following picture. Consider the cyclotron resonance condition between a proton with parallel speed $v_{\parallel}$ and a fluctuation quasi-mode $\omega(\mathbf{k})$, given by

$$\omega - k_{\parallel}\, v_{\parallel} = \Omega_p. \qquad (5)$$

This condition is represented by straight lines in the $\omega$ - $k_{\parallel}$ plane with slopes of $v_{\parallel}$ and intercepts at $(\omega, k_{\parallel}) = (\Omega_p, 0)$ as shown by the green and dashed lines in Figure 2. When one of these lines intersects a dispersion curve of the fluctuations, that proton and that fluctuation mode are cyclotron resonant. In the example of Figure 2, the blue curve is the dispersion of KAWs propagating at 89.5° to the large-scale magnetic field. The red squares indicate the potential resonances between these KAW fluctuations and protons with moderate parallel speeds both above and below $v_\parallel = 0$. This situation contrasts strongly with the standard cyclotron resonant picture of protons with quasi-parallel propagating ion cyclotron waves (Dusenbery & Hollweg 1981; Hollweg & Isenberg 2002). In that quasi-parallel case, there are no ion cyclotron modes with $\omega > \Omega_p$, and existing modes need extremely high wavenumbers to resonate with thermal protons.

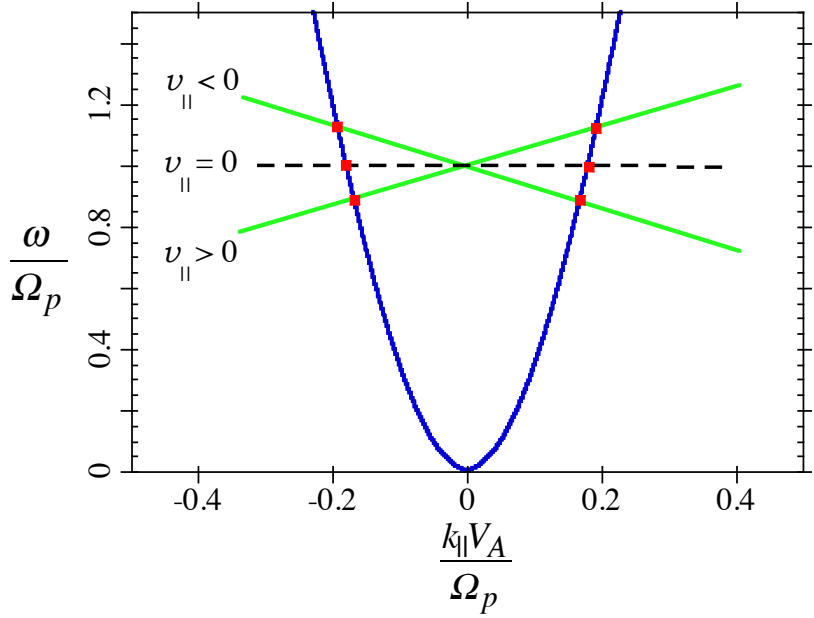

**Figure 2.** Schematic representation of proton cyclotron resonance, $\omega - k_\parallel v_\parallel = \Omega_p$ in the $(\omega, k_\parallel)$ plane. The blue curve shows KAW dispersion at 89.5°, and the straight lines correspond to parallel speeds of different protons. The red squares show the multiple resonances for each proton speed around zero.

An additional important property is evident from Figure 2. We see that when both forward- and backward-propagating KAWs ($k_\parallel \gtrless 0$) are present, any thermal proton will experience a simultaneous resonance with both modes. Thus, we expect to find additional perpendicular heating by second-order Fermi acceleration, similar to that proposed for heavy ions by Isenberg & Vasquez (2007).

## 3. Quasilinear Diffusion by a Critically-Balanced Turbulent Spectrum

In keeping with the picture presented above, we describe the QL effect that a randomly phased spectrum of quasi-modes will have on an ion distribution. For this illustrative example, we will take the QL resonances to be represented by delta functions, as derived for the case of infinite wavetrains. One can construct a QL description of the effects of turbulent quasi-modes with limited coherence by using broader functions to represent resonances of finite duration (Dupree 1966; Weinstock 1969; Lehe et al. 2009; Lynn et al. 2012). However, we will defer this extension to a future paper. We would expect broadened resonances to yield somewhat smaller diffusion from a continuous spectrum, but the qualitative behavior of the interaction and the subsequent response of the ion distributions will likely be similar.

The evolution of a gyrotropic ion distribution, $f(\upsilon_\parallel, \upsilon_\perp)$ of mass $m_i$ and charge $q_i$, is dominated by the scattering due to resonances between the ion motions and the quasi-mode fields, $\delta\mathbf{E} = \boldsymbol{\varepsilon}\,\mathrm{e}^{i(\mathbf{k}\cdot\mathbf{x}-\omega t)}$. The QL interaction is described by

$$\frac{\partial f}{\partial t} = \frac{\pi q_i^2}{m_i^2 \upsilon_\perp}\sum_n \int d^3\mathbf{k}\, G\left[\upsilon_\perp \delta(\omega - k_\parallel \upsilon_\parallel - n\Omega_i)\left|\psi_n(\mathbf{k})\right|^2 G\,f\right] \qquad (6)$$

(Kennel & Engelmann 1966; Kennel & Wong 1967a, b; Stix 1992), where the differential operator

$$G \equiv \left(1 - \frac{k_\parallel \upsilon_\parallel}{\omega}\right)\frac{\partial}{\partial \upsilon_\perp} + \frac{k_\parallel \upsilon_\parallel}{\omega}\frac{\partial}{\partial \upsilon_\parallel}. \qquad (7)$$

The delta function picks out the resonances, when the Doppler-shifted fluctuation frequency seen by an ion moving along the background magnetic field at $\upsilon_\parallel$ is equal to an integer multiple of the ion's gyrofrequency. The spectral intensity weighting function is

$$\left|\psi_n(\mathbf{k})\right|^2 = \left|\varepsilon_l\, e^{-i\varphi} J_{n-1} + \varepsilon_r\, e^{i\varphi} J_{n+1} + \frac{\upsilon_\parallel}{\upsilon_\perp}\varepsilon_z J_n\right|^2, \qquad (8)$$

where we have decomposed the electric field Fourier components into their left-hand, right-hand, and parallel polarizations with respect to the large-scale magnetic field, directed along the $z$ axis. Specifically, with respect to arbitrarily oriented $x$ and $y$ coordinates in the plane perpendicular to $z$, we have $\varepsilon_l = (\varepsilon_x + i\varepsilon_y)/2$, $\varepsilon_r = (\varepsilon_x - i\varepsilon_y)/2$, and $\varphi$ indicates the phase of the rotation for a given vector $\boldsymbol{\varepsilon}(\mathbf{k})$. The $J$'s are Bessel functions of integer order and argument $k_\perp \upsilon_\perp/\Omega_i$. The full expression (6) calls for a sum over all possible resonances, but we will consider only the $n = 0$ and $n = 1$ terms, which correspond to the Landau and fundamental ion-cyclotron resonances. Thus, for the time being we neglect the effect of anomalous resonances ($n < 0$) or higher cyclotron harmonics ($n > 1$) that require large parallel speeds for the ions. Taking the arbitrary phase such that $\mathrm{Re}(\varepsilon_y) = \mathrm{Im}(\varepsilon_x) = 0$, $\mathrm{Re}(\varepsilon_x) = E_x$, and $\mathrm{Im}(\varepsilon_y) = E_y$, expanding out the complex conjugate in (8), and then averaging over the phase gives

$$\left|\psi_n(\mathbf{k})\right|^2 = \frac{1}{4}\left[\left(E_x - E_y\right)^2 J_{n-1}{}^2 + \left(E_x + E_y\right)^2 J_{n+1}{}^2\right] + \left(\frac{\upsilon_\parallel}{\upsilon_\perp}\right)^2 E_z{}^2 J_n{}^2. \qquad (9)$$

The double application of the operator $G$ in (6) results in a diffusion equation in cylindrical velocity space

$$\frac{\partial f}{\partial t} = \frac{\partial}{\partial \upsilon_\parallel}\left[D_{\parallel,\parallel}\frac{\partial f}{\partial \upsilon_\parallel} + D_{\parallel,\perp}\frac{\partial f}{\partial \upsilon_\perp}\right] + \frac{1}{\upsilon_\perp}\frac{\partial}{\partial \upsilon_\perp}\left[\upsilon_\perp\left(D_{\parallel,\perp}\frac{\partial f}{\partial \upsilon_\parallel} + D_{\perp,\perp}\frac{\partial f}{\partial \upsilon_\perp}\right)\right]. \quad (10)$$

Performing the integral over $k_\parallel$ in (6) sets the resonant parallel wavenumbers $k_{res}$, and gives the diffusion coefficients as

$$D_{a,b} = C\int k_\perp\, dk_\perp M_{a,b}\left[\frac{\left|\psi_1(k_{res}{}^1, k_\perp)\right|^2}{\left|W(k_{res}{}^1, k_\perp) - \upsilon_\parallel\right|} + \frac{\left|\psi_0(k_{res}{}^0, k_\perp)\right|^2}{\left|W(k_{res}{}^0, k_\perp) - \upsilon_\parallel\right|}\right], \qquad (11)$$

where the first term in the square bracket comes from the $n = 1$ cyclotron resonance at $\omega - k_{res}{}^1 \upsilon_\parallel = \Omega_i$,

$$\left|\psi_1(k_{res}{}^1, k_\perp)\right|^2 = \frac{1}{4}\left[\left(E_x - E_y\right)^2 J_0{}^2 + \left(E_x + E_y\right)^2 J_2{}^2\right] + \left(\frac{\upsilon_\parallel}{\upsilon_\perp}\right)^2 E_z{}^2 J_1{}^2, \quad (12)$$

and the second term from the $n = 0$ Landau resonance at $\omega - k_{res}{}^0 \upsilon_\parallel = 0$,

$$\left|\psi_0(k_{res}{}^0, k_\perp)\right|^2 = \frac{1}{2}\left(E_x{}^2 + E_y{}^2\right)J_1{}^2 + \left(\frac{\upsilon_\parallel}{\upsilon_\perp}\right)^2 E_z{}^2 J_0{}^2. \qquad (13)$$

The denominators in the bracketed terms come from the delta function integral, where $W(k_{res}, k_\perp)$ is the parallel group speed of the quasi-mode at the resonant wavenumber, $W = \partial\omega/\partial k_\parallel$. We also define the coefficients

$$M_{\parallel,\parallel} = \left(\frac{\upsilon_\perp}{V_{ph}}\right)^2; \quad M_{\parallel,\perp} = \frac{\upsilon_\perp\left(V_{ph} - \upsilon_\parallel\right)}{V_{ph}{}^2}; \quad M_{\perp,\perp} = \left(\frac{V_{ph} - \upsilon_\parallel}{\upsilon_\perp}\right)^2; \qquad (14)$$

where $V_{ph}$ is the resonant parallel phase speed of the quasi-mode $\omega/k_{res}$, and $C = 2\pi^2 q_i^2/m_i^2$. It is clear from (14) that the Landau-resonant terms will only contribute to $D_{\parallel,\parallel}$.

We note that the transverse electric field terms in the Landau-resonant expression (13) are related to the parallel magnetic compressibility, $\delta B_z^2$, through Faraday's law. These terms correspond to a transit-time damping contribution to the Landau resonance. Thus, our expressions contain both Landau damping and transit-time damping effects.

To model the power spectrum in this illustrative example, we choose an analytic expression that is qualitatively consistent with current understanding of collisionless plasma turbulence in the upper inertial range and lower dissipation range (e.g. TenBarge & Howes (2012))

$$\left\langle \delta B^2(\mathbf{k}) \right\rangle = A \frac{(\lambda k_\perp)^{-7/3} + (\lambda k_\perp)^{-1.13}}{1 + (\lambda k_\perp)^2} S\left(\lambda k_{\parallel cb} - \lambda |k_\parallel|\right) \tag{15}$$

This spectrum combines a power law inertial range that has a one-dimensional perpendicular energy spectrum $<\delta B^2(k_\perp)> = \int <\delta B^2> \, dk_\parallel$ scaling as $k_\perp^{-5/3}$, with a steeper dissipation range dependence of $k_\perp^{-2.8}$ consistent with some observations (Alexandrova et al. 2009; Chen et al. 2010; Sahraoui et al. 2010) and simulations (Arzamasskiy et al. 2019). The smooth transition between these power laws takes place here around the proton inertial scale $\lambda k_\perp = 1$. The turbulent fluctuations are further confined to a wavevector cone, $k_\parallel \leq k_{\parallel cb}$, bounded by a surface suggested by some critical balance relations (Goldreich & Sridhar 1995; Cho & Lazarian 2004; Howes et al. 2008a; Schekochihin et al. 2009)

$$\lambda k_{\parallel cb} = (\lambda k_i)^{1/3} \frac{(\lambda k_\perp)^{2/3} + (\lambda k_\perp)^{7/3}}{1 + (\lambda k_\perp)^2} \tag{16}$$

where $k_i$ is a normalization scale at which the spectrum is assumed to be isotropic. The spectral dependence on $k_\parallel$ is taken to be a simple flat function with a sharp cutoff, given by the Heaviside step function, $S(x)$, as a simple approximation to a more realistic, gradual confinement. This model spectrum is also taken symmetric in $k_\parallel$, representing a balanced turbulent cascade, in order to investigate the simplest possibilities first.

We note that the theoretical models giving the specific critical balance boundary (16) also predict an intensity spectral index of $k_\perp^{-7/3}$, which is shallower than the $-2.8$ value used in (15). This possible inconsistency will not affect the qualitative results of our calculations. Quantitatively, the steeper spectrum used here leads to less KAW power than would have been obtained with the standard spectral slope, so our results could be viewed as a lower bound to the modeled heating.

A further idealization is found in our setting the characteristic wavenumbers in this model spectrum to scale as the proton inertial length, $\lambda$, rather than the possibly more physically motivated choice of the proton gyroradius, $\rho_p$. This idealization simply allows the model turbulent spectrum to be independent of the evolving particle distributions in this illustrative example. This choice also should not affect the qualitative results of this paper. We note that moving the spectral breakpoint in (15) to the gyroradius, for $\beta = 0.1$, would extend the inertial range where there are higher intensities, and so would lead to even stronger resonant heating than we will obtain here.

Using the KAW dispersion relations (1) and (2) with the spectral intensities given by (15) and (16), the model diffusion coefficients (11) can be calculated. The transformation of fluctuating magnetic field intensities (15) into fluctuating electric field intensities required by (12) and (13), is obtained from the Fourier transform of Faraday's law in terms of the total phase speed of the resonant quasi-mode,

$$\left\langle \delta E^2 \right\rangle = \left( \frac{k_{res} V_{ph}}{kc} \right)^2 \left\langle \delta B^2 \right\rangle. \tag{17}$$

We compute these diffusion coefficient values for $0 \leq |\upsilon_\parallel / V_A| \leq 1.28$ and $\upsilon_\perp / V_A \leq 3$, which will be applied to calculations of ion heating in the next section. The upper limits of this phase space are chosen to include the principal behavior due to both cyclotron and Landau resonances. Figures 3 and 4 show these values in units of $2\pi^2 \, \Omega_p V_A^2 (<\delta b^2>/B_o^2)$, where $(<\delta b^2>/B_o^2) = A/(\lambda^3 B_o^2)$ is the relative turbulent spectral intensity at $\lambda k_\perp = 1$, $k_\parallel \leq k_{\parallel cb}$. These figures only illustrate the diffusion coefficients up to $\upsilon_\perp / V_A = 2$ in order to show more detail, but the computations of the next section extend to $\upsilon_\perp / V_A = 3$.

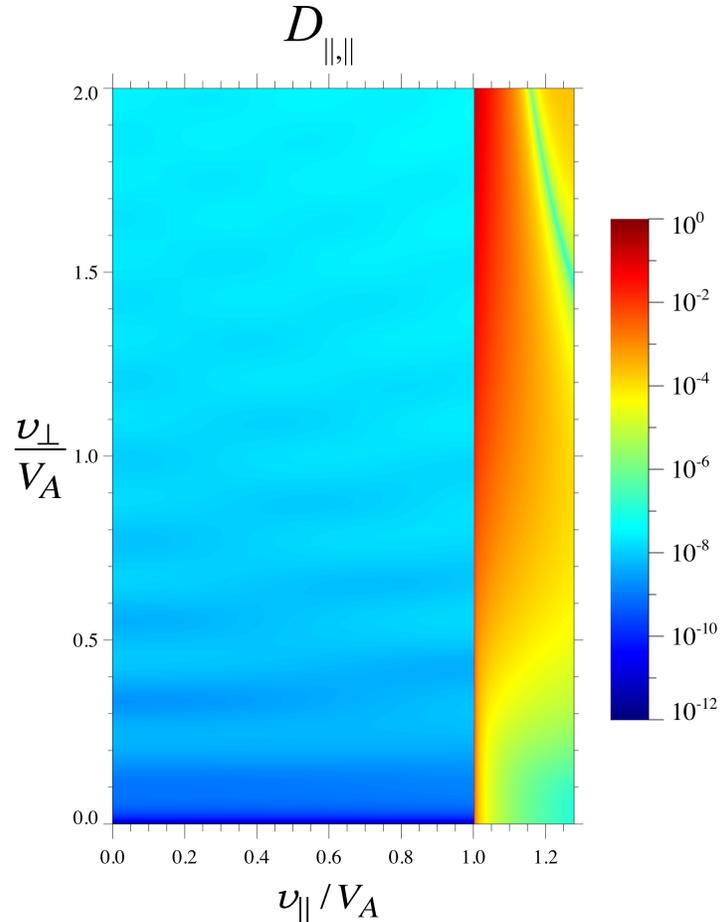

**Figure 3.** The parallel component of the diffusion coefficient (11) for protons in the KAW spectrum (15).

The diffusion coefficient in the parallel direction, $D_{\parallel,\parallel}$, shown in Figure 3, has enormous dynamic range, and is dominated by the Landau resonance for $\upsilon_{\parallel}/V_A \geq 1$, which can reach values of $10^5$ times the largest cyclotron-resonant value. This result presumably explains the many previous conclusions that the dissipation of these cascading fluctuations on ions takes place through Landau damping (Quateart 1998; Howes et al. 2008a; Schekochihin et al. 2009; Howes 2010, 2011). However, this damping is expected to be weak for our example of $\beta = 0.1$, since there are very few particles that can resonate with these fluctuations.

The other components of the diffusion coefficient are shown in Figure 4, using the same colorbar coding for both panels. These components have no Landau-resonant terms. The cross component, $D_{\parallel,\perp}$, is primarily negative, consistent with the pitch-angle-like diffusion by the backward-propagating quasi-modes. However, the oscillating nature of the Bessel functions in (8) yields a few regions where the forward-propagating resonance is locally stronger, and this gives positive values for this component. To use a logarithmic scale in the figure, we plot the absolute value and note that these positive features are confined within the blue loops on the left. The perpendicular component, $D_{\perp,\perp}$, is much larger than the other cyclotron-resonant values, indicating that the core of a proton distribution will be perpendicularly heated by this interaction.

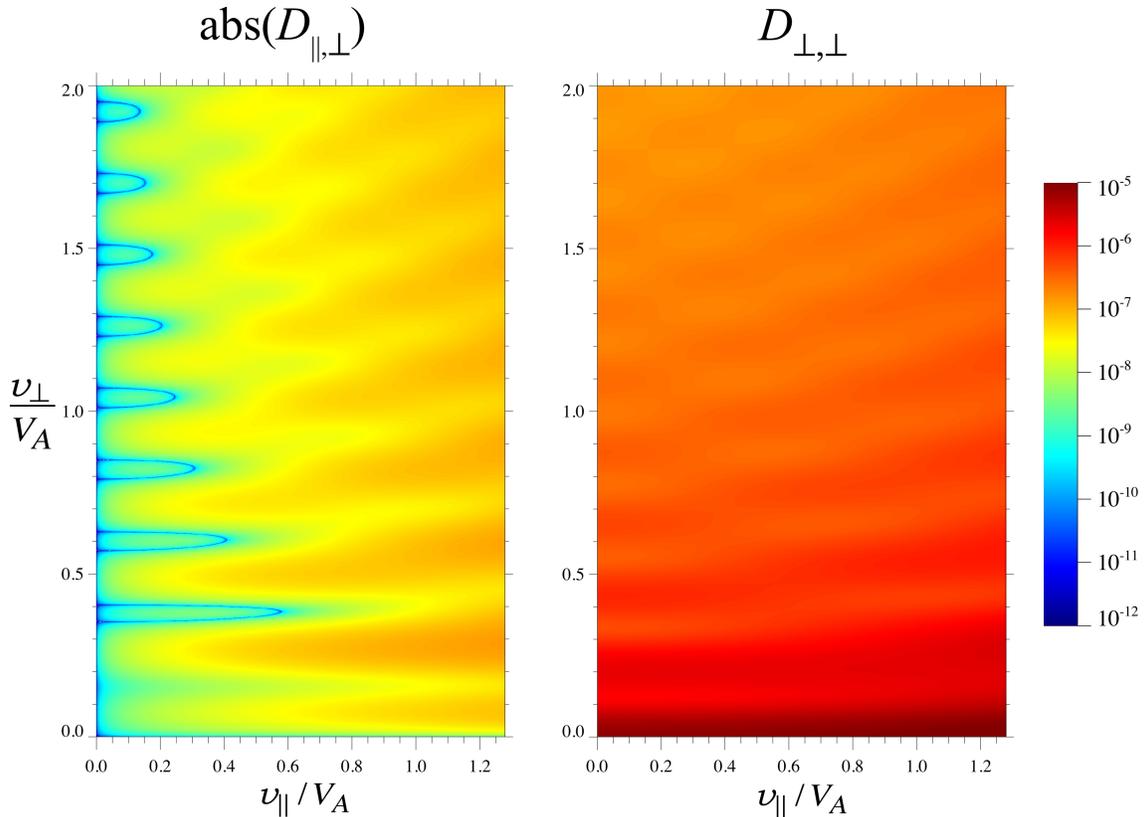

**Figure 4.** The cross and perpendicular components of the diffusion coefficient (11) for protons in the KAW spectrum (15).

These QL diffusion coefficients exhibit dependencies somewhat different from those implied by nonlinear dissipation mechanisms. For instance, the quantitative dependence of $D_{\perp,\perp}$ on the proton perpendicular speed is shown in Figure 5 for several values of the proton parallel speed. We see that the general trend is close to $D_{\perp,\perp} \sim \upsilon_{\perp}^{-1}$. This behavior contrasts with that of the perpendicular diffusion predicted by the nonlinear MMB mechanism, which calls for $D_{\perp,\perp} \sim \upsilon_{\perp}^{2}$ with an exponential cutoff at large $\upsilon_{\perp}$ (Chandran et al. 2010; Klein & Chandran 2016). In the MMB mechanism, protons with increasing perpendicular speed interact with fluctuations of decreasing $k_\perp$, whose intensities are larger. In the QL cyclotron-resonant mechanism of this paper, the dependence on $v_\perp$ primarily enters through the Bessel functions in (9), which decrease for increasing $v_\perp$.

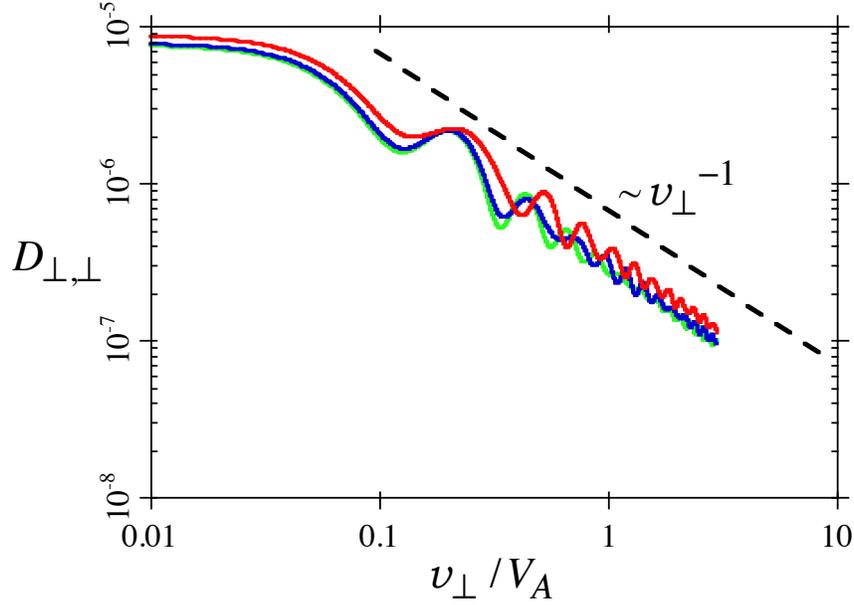

**Figure 5.** The perpendicular diffusion coefficient from (11) in the same units as Fig. 4, for three values of the proton parallel speed, $v_\parallel / V_A = 0.2$ (green), 0.4 (blue), and 0.8 (red).

## 4. Turbulent Ion Heating

### a) Protons

With these diffusion coefficients in hand, we solve the diffusion equation (10) for the gyrotropic distribution function of protons in this homogeneous system. The initial distribution function is defined on a grid in ($v_\parallel$, $v_\perp$) velocity space and the equation is discretized on that grid. We take a grid with 1280 points from $0 \le v_\parallel / V_A \le 1.28$ and 3000 points from $0 \le v_\perp / V_A \le 3$, with reflecting boundaries at the zero-speed surfaces and absorbing boundaries at the outer surfaces. The grid in $v_\perp$ is staggered to avoid the singularity in the cylindrical coordinate at $v_\perp = 0$. The time-dependent difference equation is then advanced with a flux-conservative, implicit scheme. The matrix inversions to advance the solution in time use the bi-conjugate gradient stabilized method, preconditioned by incomplete lower and upper (ILU) factorizations.

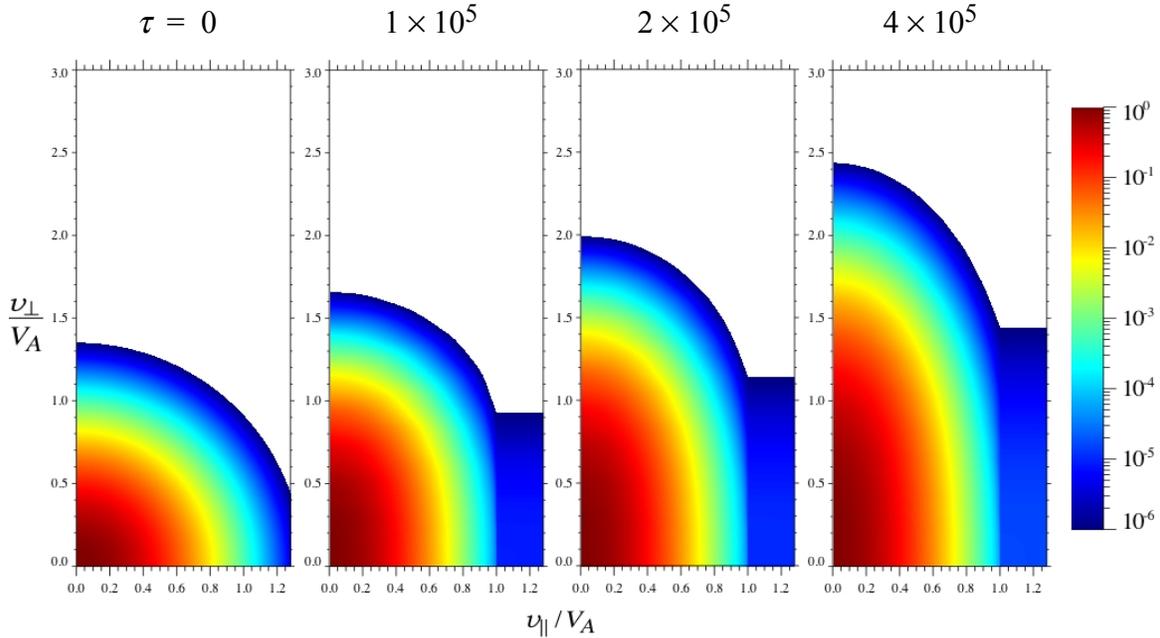

**Figure 6.** Evolving proton distribution, from (10), for four times in our computation. The distribution is normalized in each panel to the maximum value at the phase-space origin.

The computational time scale depends on the grid spacing as well as on the spectral intensity, resulting in $\tau \equiv \Omega_p t = 5 \times 10^{-8}\ (<\delta b^2>/B_o^2)^{-1}$ for this case. Computational time steps on the order of $\Delta\tau \sim 1$ conserve the total proton density to within 1.4% by the end of the computation at $\tau = 4 \times 10^5$.

Figure 6 shows the time evolution of the proton distribution, normalized to its maximum value at the origin in each panel. The computation is initialized with a Maxwellian distribution at $\beta = 0.1$, shown in the leftmost panel. As time advances, the core of the distribution is strongly energized in the perpendicular direction. There is some parallel diffusion of the protons with $\upsilon_{\|}/V_A > 1$, due to the Landau resonance, but this leakage from the core remains a secondary effect.

The total anisotropy for the computed distribution is shown in Figure 7, where we see that the protons are perpendicularly heated in a nearly linear fashion.

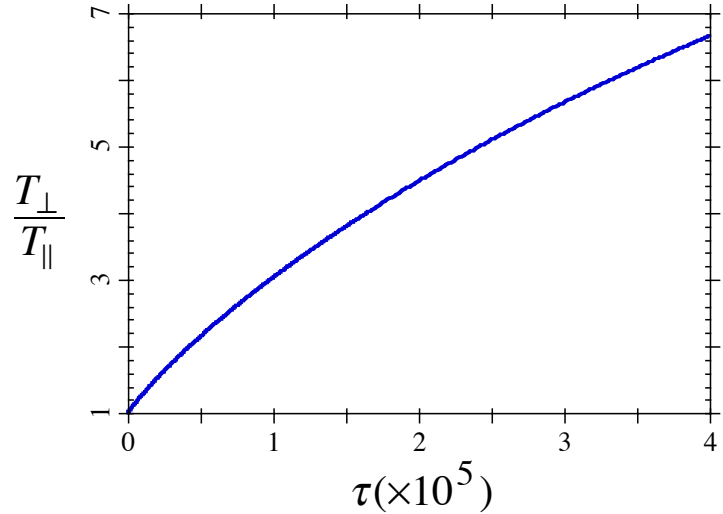

**Figure 7.** Proton anisotropy for the computation shown in Fig. 6.

To estimate the physical time scale for this heating, consider values of the relative spectral intensity, $<\delta b^2>/B_o{}^2 = 10^{-4} - 10^{-6}$, such as might be measured in the fast solar wind at a heliocentric radial position of 20 $R_s$. With these estimated intensities, the proton heating time shown in Figures 6 and 7 corresponds to several thousand to several $\times 10^5$ proton gyroperiods. This is a reasonable time scale for an efficient QL process. Taking the proton gyrofrequency near 20 $R_s$ to be $\Omega_p = 20$ s$^{-1}$ gives an estimated time between 10 and 1000 seconds for the proton heating shown in these figures.

This example implies that the resonant cyclotron heating of solar wind protons by turbulent KAWs can be very fast, at least for plasma parameters near 20 $R_s$, where $\beta \sim$ 0.1. The coupled dissipation of the KAW fluctuations is not included here, but would necessarily be limited by the turbulent cascade rate to continually replenish the resonant fluctuations.

### b) Alpha particles

In addition to preferential energization in the perpendicular directions, the kinetic mechanism(s) responsible for solar wind heating must also supply more thermal energy to heavy ions relative to the proton population. To test this property within the context of our example, we also compute the response of an alpha particle distribution to the same turbulent spectrum of KAWs.

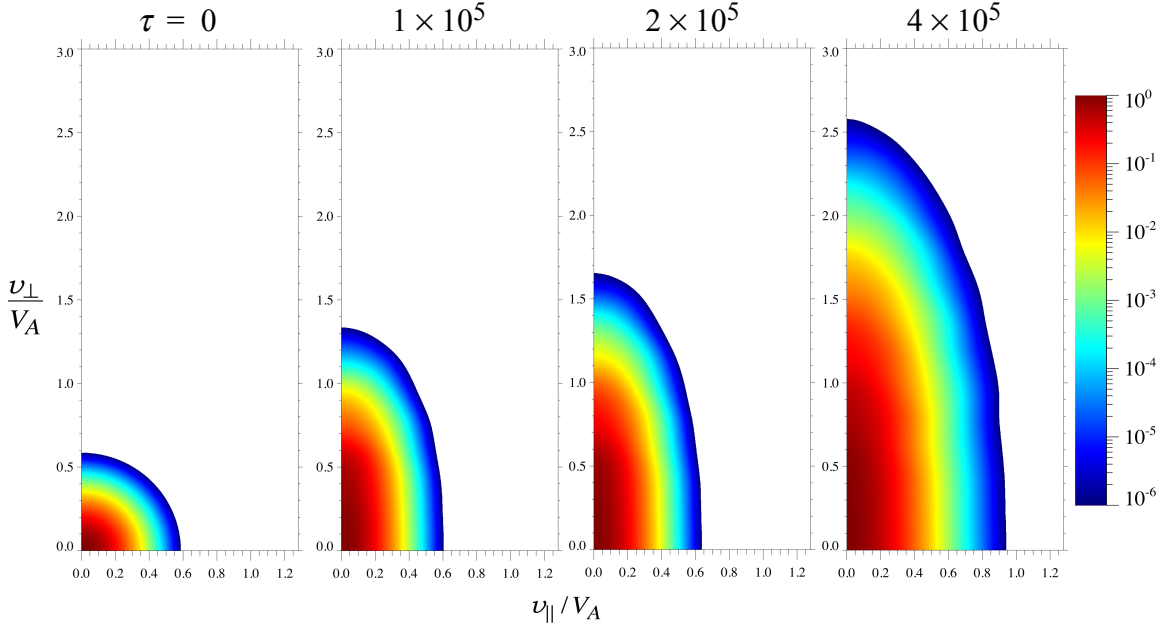

**Figure 8.** Evolving alpha particle distribution, from (10), for four times in our computation. The distribution is normalized in each panel to the maximum value at the phase-space origin.

The QL interaction with alpha particles is only slightly modified from the proton case presented above. In equation (6), the resonance condition is shifted to the lower gyrofrequency, $\Omega_\alpha = \Omega_p/2$, and the visualization of this condition in Figure 2 similarly shifts the straight lines to the lower intercept at $(\omega, k_\parallel) = (\Omega_\alpha, 0)$. This shift to lower frequency corresponds to resonance with higher intensity fluctuations in the power-law spectrum (15). At the same time the overall coefficient, $C$ in equation (11), is reduced by a factor of 4. The resulting diffusion coefficients (not shown) have a structure similar to the proton coefficients in Figures 3 and 4, with somewhat higher values.

We perform the same computation of QL diffusion on alpha particles as was shown in Figure 6 for protons, starting with a $\beta = 0.1$ Maxwellian distribution and using the same KAW turbulent spectrum with the new alpha particle diffusion coefficients. The results are shown in Figure 8 for the same duration, $\tau \rightarrow 4 \times 10^5$, corresponding to physical times of 10 - 1000 seconds near 20 $R_s$ in the solar wind.

We see that the alpha particles attain perpendicular thermal speeds comparable to, but slightly larger than, the protons during the same time period. This heating is consistent with the greater than mass proportional perpendicular temperatures of heavy ions that are frequently observed in the fast solar wind. The small amount of parallel heating, when acting on the reduced parallel thermal spread of the initial alpha particle distribution, leads to large anisotropies, as shown in Figure 9. It is clear from Figure 8 that the alpha particle perpendicular temperature continues to increase with time, and that the anisotropy in Figure 9 only levels off due to the slower parallel heating that also takes place.

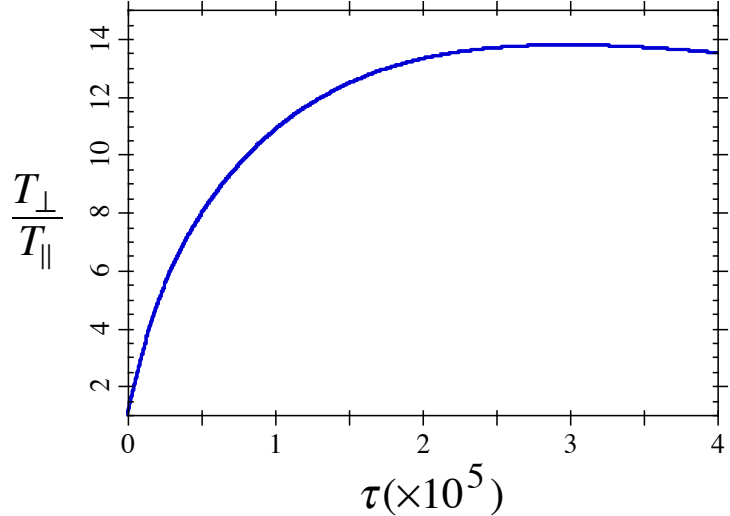

**Figure 9.** Alpha particle anisotropy for the computation shown in Fig. 8.

## 5. Discussion

These examples, though idealized, demonstrate how solar wind ions may be heated in the perpendicular direction by dissipation of collisionless plasma turbulence, consistent with coronal and *in situ* observations. In this paper, we considered a homogeneous collisionless plasma with $\beta = 0.1$ interacting with a steady spectrum of highly oblique KAWs. This KAW spectrum was chosen to represent a reasonable model turbulent fluctuation spectrum in the range around the proton inertial scale. We found that ion heating by the resonant cyclotron interaction with these KAWs yields preferential energization in the perpendicular direction, and greater than mass proportional heating on heavy ions. This heating may be very efficient, and our idealized example produces significant effects in a very short time.

It is therefore possible that the kinetic mechanism that heats solar wind ions, and perhaps drives the fast solar wind in the corona, could turn out to be the familiar resonant cyclotron interaction, acting to dissipate turbulently generated oblique KAWs. However, further investigation and application of a number of physically based improvements to the work presented here are required before such a claim can be made. We discuss several of these improvements below.

A major idealization of this paper is the use of KAW dispersion relations derived from two-fluid analysis. A two-fluid system implicitly assumes that the ions and electrons maintain Maxwellian distributions, and the kinetic effects of instability or damping on the fluctuations are neglected. This approximation should be reasonable at frequencies and wavenumbers that are not too close to the particle resonances. For the fluctuations considered here, the two-fluid analysis indicates likely valid solutions both above and below $\omega = \Omega_p$. Near the resonances one may expect some distortion of the expressions and, in principle, an increase in the damping rate, derivable from a kinetic analysis. If the damping is strong, however, the ion distribution will not remain close to a Maxwellian and the standard kinetic computations will still not be valid. What is needed to improve the example of this paper is a coupled computation accounting for the simultaneous evolution of both the kinetic KAW dispersion and the ion distributions. We will construct such a coupled computation in our next work.

A dispersion relation that evolves with the changing proton distribution may allow this efficient mechanism to operate at lower plasma $\beta$, such as expected in the solar corona. Perpendicularly enhanced protons will lead to an increased phase speed for the fluctuations, which may extend the frequency range of these oblique KAWs. For instance, the isotropic two-fluid expressions in this paper do not yield solutions that pass through $\Omega_p$ when the plasma $\beta$ is lowered to 0.01. We will determine if a self-consistent increase in the proton anisotropy can remove this limitation.

A further physical improvement to this work will follow from modifying the QL diffusion expressions to allow broadening of the sharp, delta-function interactions used here. In taking the KAW spectrum in equations (15) and (16) to represent strong, critically balanced turbulence, we should account for their turbulent decorrelation in the QL interaction integrals (Dupree 1966; Weinstock 1969; Lehe et al. 2009; Lynn et al. 2012). It will be interesting to see how these heating rates are modified by such resonance broadening.

In the case where the only interaction being modeled is this QL dissipation of KAWs, one would expect that a steady spectrum representing driven turbulence would produce large ion anisotropies, such as those obtained here. However, large anisotropies will trigger the well-known resonant cyclotron anisotropy instability (Isenberg et al. 2019), which should operate with comparable efficiency to scatter the ions into the parallel direction. This scattering is accompanied by QL generation of parallel-propagating ion-cyclotron waves, which are often observed in the solar wind. A realistic model of strong perpendicular ion heating would need to include the limiting effects of this interaction.

On the other hand, there appears to be no inherent limit to the total ion heating from this mechanism as long as the turbulent fluctuations are maintained. This situation stands in contrast to the proposed nonlinear mechanisms of magnetic-moment-breaking or small-scale reconnection that become less efficient for large gyroradii. Thus, this mechanism may also provide some of the energization needed to produce the observed suprathermal ion population in the solar wind (Fisk & Gloeckler 2007; Dayeh et al. 2009).

Finally, if the heating process presented here remains a realistic candidate for ion heating in the solar wind, its efficiency needs to be compared to an estimated turbulent cascade rate. In this paper, we have assumed a steady spectrum of fluctuations, but the dissipative heating will ultimately depend on the turbulent cascade to replenish the fluctuation power in the dissipation range of the spectrum. In a true steady state system, the dissipation rate should be comparable to the cascade rate, and this relation would determine the net ion heating in the solar wind.

## 6. Summary and Conclusions

The kinetic mechanism by which plasma turbulence in the solar corona and inner heliosphere may dissipate to yield the observed perpendicular ion heating in the solar wind has been a puzzle for several decades. We suggest that, if the turbulence at ion scales can be characterized as highly oblique KAWs, these fluctuations can dissipate through the quasilinear resonant cyclotron interaction and produce strong perpendicular heating of thermal ions. We demonstrate this process with an idealized example in a homogeneous plasma with $\beta = 0.1$. We take the turbulence to be represented by a steady, critically balanced spectrum of KAWs, described by a two-fluid dispersion relation. In this case, we show that both thermal protons and alpha particles are heated to substantial perpendicular temperatures in a short time. Both ion species attain comparable perpendicular thermal speeds, consistent with the observations of generally mass proportional perpendicular temperatures in the fast solar wind. Thus, this quasilinear process of resonant cyclotron heating may ultimately explain the perpendicular ion heating that drives the solar wind.

**Acknowledgements.** The authors are grateful for valuable conversations with L. Arzamasskiy, K. G. Klein, R. Kumar, M. W. Kunz, C. W. Smith, J. M. TenBarge, and D. Verscharen. We also thank the anonymous referee for comments that led to improvements in this paper. Computations were performed on Trillian, a Cray XE6m-200 supercomputer at UNH supported by the NSF MRI program under grant PHY-1229408. This work was also supported in part by NASA grants 80NSSC17K0009 and 80NSSC18K1215.

**References**

Alexandrova, O., Saur, J., Lacombe, C., et al. 2009, Phys. Rev. Lett., 103, 165003
Arzamasskiy, L., Kunz, M. W., Chandran, B. D. G., & Quataert, E. 2019, Astrophys. J., 879, 53
Bale, S. D., Kellogg, P. J., Mozer, F. S., Horbury, T. S., & Rème, H. 2005, Phys. Rev. Lett., 94, 215002
Bian, N. H., Kontar, E. P., & Brown, J. C. 2010, Astron. Astrophys., 519, A114
Bochsler, P., Geiss, J., & Joos, R. 1985, J. Geophys. Res., 90, 10779
Boldyrev, S., & Perez, J. C. 2012, Astrophys. J., 758, L44
Boldyrev, S., Horaites, K., Xia, Q., & Perez, J. C. 2013, Astrophys. J., 777, 41
Chandran, B. D. G., Li, B., Rogers, B. N., Quataert, E., & Germaschewski, K. 2010, Astrophys. J., 720, 503
Chen, C. H. K., Boldyrev, S., Xia, Q., & Perez, J. C. 2013, Phys. Rev. Lett., 110, 225002

Chen, C. H. K., Horbury, T. S., Schekochihin, A. A., et al. 2010, Phys. Rev. Lett., 104, 255002

Chen, L., Lin, Z., & White, R. 2001, Phys. Plasmas, 8, 4713

Cho, J., & Vishniac, E. T. 2000, Astrophys. J., 539, 273

Cho, J., & Lazarian, A. 2004, Astrophys. J., 615, L41

Cho, J., Lazarian, A., & Vishniac, E. T. 2002, Astrophys. J., 564, 291

Collier, M. R., Hamilton, D. C., Gloeckler, G., Bochsler, P., & Sheldon, R. B. 1996, Geophys. Res. Lett., 23, 1191

Cranmer, S. R. 2009, Living Rev. Solar Phys., 6, 3

Cranmer, S. R., & van Ballegooijen, A. A. 2003, Astrophys. J., 594, 573

———. 2005, Astrophys. J. Suppl., 156, 265

Cranmer, S. R., van Ballegooijen, A. A., & Edgar, R. J. 2007, Astrophys. J. Suppl., 171, 520

Cranmer, S. R., Panasyuk, A. V., & Kohl, J. L. 2008, Astrophys. J., 678, 1480

Dayeh, M. A., Desai, M. I., Dwyer, J. R., et al. 2009, Astrophys. J., 693, 1588

Dmitruk, P., Matthaeus, W. H., Milano, L. J., & Oughton, S. 2001, Phys. Plasmas, 8, 2377

Dupree, T. H. 1966, Phys. Fluids, 9, 1733

Dusenbery, P. B., & Hollweg, J. V. 1981, J. Geophys. Res., 86, 153

Esser, R., Fineschi, S., Dobrzycka, D., et al. 1999, Astrophys. J., 510, L63

Fisk, L. A., & Gloeckler, G. 2007, Space Sci. Rev., 130, 153

Gary, S. P. 1986, J. Plasma Phys., 35, 431

Gary, S. P., & Borovsky, J. E. 2004, J. Geophys. Res., 109, 6105

Gary, S. P., & Nishimura, K. 2004, J. Geophys. Res., 109, 2109

Gary, S. P., & Borovsky, J. E. 2008, J. Geophys. Res., 113, A12104

Goldreich, P., & Sridhar, S. 1995, Astrophys. J., 438, 763

Grošelj, D., Mallet, A., Loureiro, N. F., & Jenko, F. 2018a, Phys. Rev. Lett., 120, 105101

Grošelj, D., Chen, C. H. K., Mallet, A., et al. 2018b, arXiv:1806.05741

He, J., Tu, C.-Y., Marsch, E., & Yao, S. 2012, Astrophys. J., 745, L8

He, J., Marsch, E., Tu, C.-Y., Yao, S., & Tian, H. 2011, Astrophys. J., 731, 85

Hefti, S., Grünwaldt, H., Ipavich, F. M., et al. 1998, J. Geophys. Res., 103, 29697

Hellinger, P., Trávníček, P., Stverak, S., Matteini, L., & Velli, M. 2013, J. Geophys. Res., 118, 1351

Higdon, J. C. 1984, Astrophys. J., 285, 109

Hollweg, J. V. 1999, J. Geophys. Res., 104, 14,811

Hollweg, J. V., & Isenberg, P. A. 2002, J. Geophys. Res., 107, 1147

Howes, G. G. 2010, Mon. Not. R. Astron. Soc., 409, L104

Howes, G. G. 2011, Astrophys. J., 738, 40

Howes, G. G., & Quataert, E. 2010, Astrophys. J., 709, L49

Howes, G. G., Klein, K. G., & TenBarge, J. M. 2014, arXiv:1404.2913

Howes, G. G., Cowley, S. C., Dorland, W., et al. 2008a, J. Geophys. Res., 113, A05103

Howes, G. G., Dorland, W., Cowley, S. C., et al. 2008b, Phys. Rev. Lett., 100, 065004

Isenberg, P. A., & Vasquez, B. J. 2007, Astrophys. J., 668, 546

———. 2011, Astrophys. J., 731, 88

———. 2015, Astrophys. J., 808, 119

Isenberg, P. A., Vasquez, B. J., & Hollweg, J. V. 2019, Astrophys. J., 870, 119

Johnson, J. R., & Cheng, C. Z. 2001, Geophys. Res. Lett., 28, 4421

Karimabadi, H., Roytershteyn, V., Wan, M., et al. 2013, Phys. Plasmas, 20, 012303

Kennel, C. F., & Engelmann, F. 1966, Phys. Fluids, 9, 2377

Kennel, C. F., & Wong, H. V. 1967a, J. Plasma Phys., 1, 75

——. 1967b, J. Plasma Phys., 1, 81

Klein, K. G., & Chandran, B. D. G. 2016, Astrophys. J., 820, 47

Klein, K. G., Howes, G. G., TenBarge, J. M., & Podesta, J. J. 2014, Astrophys. J., 785, 138

Klein, K. G., Howes, G. G., TenBarge, J. M., et al. 2012, Astrophys. J., 755, 159

Kohl, J. L., Noci, G., Antonucci, E., et al. 1998, Astrophys. J., 501, L127

Leamon, R. J., Smith, C. W., Ness, N. F., & Matthaeus, W. H. 1998, J. Geophys. Res., 103, 4775

Leamon, R. J., Smith, C. W., Ness, N. F., & Wong, H. K. 1999, J. Geophys. Res., 104, 22,331

Leamon, R. J., Matthaeus, W. H., Smith, C. W., et al. 2000, Astrophys. J., 537, 1054

Lehe, R., Parrish, I. J., & Quataert, E. 2009, Astrophys. J., 707, 404

Lionello, R., Velli, M., Downs, C., et al. 2014, Astrophys. J., 784, 120

Lynn, J. W., Parrish, I. J., Quataert, E., & Chandran, B. D. G. 2012, Astrophys. J., 758, 78

Lysak, R. L., & Lotko, W. 1996, J. Geophys. Res., 101,

Marsch, E. 1991, in Physics of the Inner Heliosphere. 2. Particles, Waves and Turbulence, ed. R. Schwenn, & E. Marsch (Berlin: Springer-Verlag), 45

——. 2006, Living Rev. Solar Phys., 3, 1

——. 2012, Space Sci. Rev., 172, 23

Matthaeus, W. H., & Lamkin, S. L. 1986, Phys. Fluids, 29, 2513

Matthaeus, W. H., & Velli, M. 2011, Space Sci. Rev., 160, 145

Matthaeus, W. H., Zank, G. P., Oughton, S., Mullan, D. J., & Dmitruk, P. 1999, Astrophys. J., 523, L93

Matthaeus, W. H., Oughton, S., Osman, K. T., et al. 2014, Astrophys. J., 790, 155

McChesney, J. M., Stern, R. A., & Bellan, P. M. 1987, Phys. Rev. Lett., 59, 1436

Montgomery, D., & Turner, L. 1981, Phys. Fluids, 24, 825

Narita, Y., Gary, S. P., Saito, S., Glassmeier, K.-H., & Motschmann, U. 2011, Geophys. Res. Lett., 38, L05101

Oughton, S., Priest, E. R., & Matthaeus, W. H. 1994, J. Fluid Mech., 280, 95

Perez, J. C., & Chandran, B. D. G. 2013, Astrophys. J., 776, 124

Podesta, J. J. 2012, J. Geophys. Res., 117, A07101

——. 2013, Solar Phys., 286, 529

Podesta, J. J., & Gary, S. P. 2011, Astrophys. J., 734, 15

Podesta, J. J., & TenBarge, J. M. 2012, J. Geophys. Res., 117, A10106

Quataert, E. 1998, Astrophys. J., 500, 978

Rappazzo, F., Velli, M., Einaudi, G., & Dahlburg, R. B. 2008, Astrophys. J., 677, 1348

Roberts, O. W., Li, X., & Li, B. 2013, Astrophys. J., 769, 58

Sahraoui, F., Belmont, G., & Goldstein, M. L. 2012, Astrophys. J., 748, 100

Sahraoui, F., Goldstein, M. L., Robert, P., & Khotyaintsev, Y. V. 2009, Phys. Rev. Lett., 102, 231102

Sahraoui, F., Goldstein, M. L., Belmont, G., Canu, P., & Rezeau, L. 2010, Phys. Rev. Lett., 105, 131101

Salem, C. S., Howes, G. G., Sundkvist, D., et al. 2012, Astrophys. J., 745, L9

Schekochihin, A. A., Cowley, S. C., Dorland, W., et al. 2009, Astrophys. J. Suppl., 182, 310

Schwartz, S. J., & Marsch, E. 1983, J. Geophys. Res., 88, 9919

Shebalin, J. V., Matthaeus, W. H., & Montgomery, D. 1983, J. Plasma Phys., 29, 525

Stix, T. H. 1992, Waves in Plasmas (New York: Am. Inst. of Phys.)

TenBarge, J. M., & Howes, G. G. 2012, Phys. Plasmas, 19, 055901

Tracy, P. J., Kasper, J. C., Raines, J. M., et al. 2016, Phys. Rev. Lett., 116, 255101

Verdini, A., & Velli, M. 2007, Astrophys. J., 662, 669

Verdini, A., Velli, M., & Buchlin, E. 2009, Astrophys. J., 700, L39

Verdini, A., Velli, M., Matthaeus, W. H., Oughton, S., & Dmitruk, P. 2010, Astrophys. J., 708, L116

von Steiger, R., Geiss, J., Gloeckler, G., & Galvin, A. B. 1995, Space Sci. Rev., 72, 71

Weinstock, J. 1969, Phys. Fluids, 12, 1045

White, R., Chen, L., & Lin, Z. 2002, Phys. Plasmas, 9, 1890

Wu, H., Verscharen, D., Wicks, R. T., et al. 2019, Astrophys. J., 870, 106

Zhao, J. S., Voitenko, Y., Yu, M. Y., Lu, J. Y., & Wu, D. J. 2014, Astrophys. J., 793, 107